\documentclass[12pt,a4paper]{article}

\usepackage[authoryear]{natbib}

\usepackage[]{graphicx}
\usepackage{enumerate}
\usepackage{amsmath}
\usepackage{amssymb}
\usepackage{url}

\usepackage{pdflscape}
\usepackage{longtable}

\newcommand{\citepp}[1]{\citep{#1}}
\newcommand{\citett}[1]{\citet{#1}}

\newcommand{\eref}[1]{(\ref{#1})}

\newcommand{\E}{\textrm{E}}
\newcommand{\N}{\textrm{N}}
\newcommand{\dd}{\textrm{d}}
\newcommand{\Jyvaskyla}{Jyv{\"a}skyl{\"a}}

\begin{document}

\title{Prioritizing covariates in the planning of future studies in the meta-analytic framework}

\author{Juha Karvanen$^1$ and Mikko J. Sillanp{\"a}{\"a}$^2$\\
~\\
$^1$Department of Mathematics and Statistics,\\
University of \Jyvaskyla,\\
\Jyvaskyla, Finland\\
juha.t.karvanen@jyu.fi\\
~\\
$^2$Department of Mathematical Sciences  and Biocenter Oulu,\\ 
University of Oulu,\\
Oulu, Finland}

\maketitle

\begin{abstract}
 Science can be seen as a sequential process where each new study augments evidence to the existing knowledge. To have the best prospects to make an impact in this process, a new study should be designed optimally taking into account the previous studies and other prior information. We propose a formal approach for the covariate prioritization, i.e., the decision about the covariates to be measured in a new study. The decision criteria can be based on conditional power, change of the p-value, change in lower confidence limit, Kullback-Leibler divergence, Bayes factors, Bayesian false discovery rate or difference between prior and posterior expectation. The criteria can be also used for decisions on the sample size. As an illustration, we consider covariate prioritization based on genome-wide association studies for C-reactive protein levels and make suggestions on the genes to be studied further.\\
 ~\\
 keywords:
design;  evidence-based medicine; meta-analysis; power; scientific method
\end{abstract}

\section{Introduction}
The efficient allocation of resources is desirable in all areas of society, including research. Statistical methods have frequently been  applied to optimize the design in experimental and observational studies. We extend the scope of optimality considerations from a single study to the research as an ongoing cumulative process where each new study augments evidence to the existing knowledge. Here optimality means that the new study will increase our knowledge as much as possible. We consider covariate prioritization, i.e., the decision which covariates should be measured in a new study and introduce tools that may help to answer the question ``What should be studied or measured next?''  The costs and the practical limitations are not directly examined but the number of covariates in the new study is assumed to be restricted.

As a motivating example, we consider replication studies for genome wide association studies (GWAS) \citepp{mccarthy2008genome,johnson2009open}  where the aim is to find genes that are associated with a disease outcome or a phenotype. The task is to select genes for an additional replication study on the basis of a published meta-analysis of the original GWAS and earlier replication studies. Conceptually, genes (or covariates in the general case)  can be divided into three categories: 
\begin{enumerate}
 \item In category I, there is already convincing evidence that the covariate is an important predictor for the outcome. Further studies are not needed to increase this evidence but there may still be good reasons to include the covariate in new studies. For instance, known or suspected confounders should be included as well as covariates needed to study interactions and subgroup effects.
 \item  In category II, the covariate potentially has an effect of clinical importance but the sample sizes in the earlier studies are inadequate for making conclusive judgements. Further studies could create new knowledge that will allow reclassify a covariate in category II  into category I or III.  
 \item In category III, the prior knowledge and existing studies show convincing evidence of the absence of a clinically significant effect. New studies with affordable sample sizes are not expected to change the conclusions.
\end{enumerate}
 The categorization involves some level of subjectivity because there are several ways to quantify the limits between the categories. It is assumed that issues related to multiple testing are taken into account.
 
We suggest that replication studies, i.e. studies where existing well-specified scientific questions are considered, should concentrate on the covariates in category II. We restrict ourselves to setups where the covariates are directly comparable and their importance can be measured by the effect size.  The evidence from the new study is combined with the existing evidence using meta-analysis. In this setup, the scientific impact of a new study may be conceptually defined as a function of the importance of the problem and the change in the knowledge due to the new study. The scientific impact is high if the study creates a lot of new knowledge on a very important question. If the problem is not important or the study creates only a small amount of new knowledge, the scientific impact is low. 

We study various approaches to the covariate prioritization in the design phase of  replication studies. According to our knowledge, the problem has not been considered systematically earlier. The closest related work by \citett{nikolakopoulou2015planning} considers the selection of study designs, treatments and sample sizes in a network meta-analysis.
Other related works are by 
\citett{Sutton:evidencebased} and 
\citett{Roloff:planning} but they consider only sample size determination, not covariate prioritization. The connections to meta-analysis, multi-stage study design and covariate selection in statistical modeling are briefly discussed in the next. 

There are only a few examples where meta-analysis has been used  to guide the study design.  
\citett{Sutton:evidencebased} used the results of a Bayesian meta-analysis as a starting point and considered sample size determination for a new study using conditional power in the updated meta-analysis as the design criterion. They found that conditional power can be highly dependent on the statistical model used in meta-analysis and even very large studies may have only a small impact on the population level estimates when there is a considerable heterogeneity between the studies. 
\citett{Roloff:planning} proposed  a conceptually similar approach that optimizes conditional power in the frequentist setup and avoids Markov Chain Monte Carlo (MCMC) computations.  The related questions include the sufficiency of the evidence from cumulative meta-analyses \citepp{ferreira2012further,langan2012graphical,Wetterslev:trial_sequential} and the quantification of the additional sample size needed due to the study heterogeneity \citepp{Wetterslev:information_size}. Conditional power has been also used to inform the design of future clinical trials in network meta-analysis \citepp{nikolakopoulou2014using}. Recently, \citett{nikolakopoulou2015planning} proposed a methodology to prioritize further research taking into account the findings of a network meta-analysis model within the context of total gain in precision.
Methods to find out whether a meta-analysis should be updated with the latest studies have been also proposed \citepp{Barrowman:identifying,Sutton:encouraging}. From a different perspective, a decision theoretic framework for the adaptation of health-care technologies has been proposed \citepp{Claxton:rational,Claxton:pilot}. The framework provides a rational way to compare the costs and benefits of additional data collection with the cost of the uncertainty.

A study designed using the meta-analysis can be seen as a two-stage study where the `first stage' has already been  published as a meta-analysis and we are planning the `second stage'. An important difference is that in multi-stage studies, the new measurements are made for individuals already in the study whereas in our situation, completely new subjects are sampled for the new study. Similar setting is present in the phase II futility study design of phase III clinical trials \citepp{levin2005utility}.  

Variable selection is a widely studied topic in statistical modeling. Commonly used variable selection criteria: Bayesian information criterion (BIC) \citepp{schwarz1978estimating}, Akaike information criterion (AIC) \citepp{akaike1974new} and least absolute shrinkage and selection operator (LASSO) \citepp{tibshirani1996regression} are based on likelihood or individual level data which we do not have available for the new study at the design phase. Differently to model selection, our aim is to select the covariates before the new study is carried out. In genetics, for instance, replication studies \citepp{little2009strengthening,sillanpaa2004replication,karvanen2009impact} have been commonly used to confirm or discard the findings from GWAS and linkage studies \citepp{ott1999analysis}. However, the choice of the data / pedigree to replicate has been based on expert opinions and expected lod score \citepp{ott1999analysis} thresholds instead of the explicit consideration of expected scientific impact at marker level. Haplotype tagging single-nucleotide polymorphism (SNP) selection exercises operate at marker level but have different targets \citepp{meng2003selection,lin2004finding}. 

The detailed definition of the problem is given in Section~\ref{sec:probledefinition}. In Section~\ref{sec:scientificimpact}, seven possible criteria for the covariate prioritization are presented. The formulation and use of these criteria for this purpose has not been proposed earlier although their building blocks may not be new (e.g., conditional power has been proposed for sample size determination).
The criteria are applied in practice in Section~\ref{sec:illustration} where a meta-analysis of GWAS for C-reactive protein (CRP) levels are considered from the viewpoint of expected scientific impact. Conclusions are presented in Section~\ref{sec:discussion}.

\section{Problem definition} \label{sec:probledefinition}
\subsection{Covariate prioritization in fixed and random effects meta-analysis}
Consider a parametric statistical model where regression coefficient $\beta_k$, $k=1,\ldots,K$ describes the individual level effect of the covariate $k$ on the outcome. A sequence of studies is carried out and after each stage meta-analysis is applied to update our knowledge on $\beta_k$.  The data collected at stage $j$ is denoted by $D_j$ and may comprise one or more studies. Under the Bayesian framework, assume that all existing information on $\beta_k$ after stage 1 can be expressed in the form of a distribution
\begin{equation} \label{eq:prior}
\beta_{k} \vert D_1 \sim \N(\mu_{1k},\sigma_{1k}^{2}),
\end{equation}
where $\mu_{1k}$ and $\sigma_{1k}^{2}$ are the expected value and the variance of $\beta_{k}$. If the prior distribution for $\beta_k$ before any data are collected is uninformative, the normality assumption can be usually justified either directly or after a suitable transformation of the parameter.  

After collecting new data in stage 2, the cumulative information on $\beta_k$ can be expressed in the form of a distribution
\begin{equation} \label{eq:posterior}
\beta_{k} \vert D_1,D_2 \sim \N(\mu_{2k},\sigma_{2k}^{2}),
\end{equation}
where $D_2$ represents the data from stage 2 and $\mu_{2k}$ and $\sigma_{2k}^{2}$ are the expected value and the variance of $\beta_k$. In other words, the studies in stage 2 updated our knowledge from $\N(\mu_{1k},\sigma_{1k}^{2})$ to $\N(\mu_{2k},\sigma_{2k}^{2})$. The scientific impact of this update consist of three components: the improvement of the precision $\sigma_{1k}^{2} \rightarrow \sigma_{2k}^{2}$, the change of the mean $\mu_{1k} \rightarrow \mu_{2k}$ and the clinical/practical importance of covariate $k$ as a predictor. It is assumed that the uncertainty on $\beta_k$ is only due to our lack of knowledge and the variance will approach zero if more studies are conducted.

The problem of interest has three aspects:
\begin{enumerate}
 \item How to measure the (realized) scientific impact after new study has been carried out, i.e. when $\mu_{2k}$ and $\sigma_{2k}^{2}$ are known.
 \item How to estimate the expected scientific impact before the new study, i.e. when $\mu_{1k}$ and $\sigma_{1k}^{2}$ are known but $\mu_{2k}$ and $\sigma_{2k}^{2}$ are unknown.
\item How to use the expected scientific impact to guide covariate prioritization.
\end{enumerate}

The problem formulation can be related to the meta-analysis in several ways. In fixed effects meta-analysis, the studies are assumed to be homogeneous and $\beta_k$ represents the true effect common for all studies. An estimate from a single study differs from $\beta_k$ only because of sampling error. It is assumed here that there is no publication bias in the meta-analysis. The point estimates and standard errors from a frequentist meta-analysis are interpreted in a Bayesian way \citepp{schmid2004bayesian} by assuming the expected value of  distribution~\eref{eq:prior} equals the point estimate and the standard deviation of distribution~\eref{eq:prior} equals the standard error. When estimating the expected scientific impact before the new study, variance $\sigma_{2k}^{2}$ can be usually reliably approximated on the basis of the variance $\sigma_{1k}^{2}$ and the planned sample size. Parameter $\mu_{2k}$ cannot be approximated in a similar manner but its expectation can be assumed to equal $\mu_{1k}$.

Often a more realistic starting point is a random effects meta-analysis where the studies are assumed to be heterogenous. Then the true effect in study $j$ can be modeled with a hierarchical model 
\begin{equation*}
 \beta_{kj} \sim \N(\beta_k,\gamma^2),
\end{equation*}
where  $\gamma^2$ describes the variability between the studies. Now estimate $\hat{\beta}_{kj}$ in study $j$ differs then from $\beta_k$ because of both study heterogeneity and sampling error. The study heterogeneity limits the impact a single study may have no matter how large the sample size is \citepp{Sutton:evidencebased}. However, if the interest still lies on parameter $\beta_k$, adding a layer for study heterogeneity does not change the problem definition: before the new study our information is described as $\beta_{k} \vert D_1 \sim \N(\mu_{1k},\sigma_{1k}^{2})$ and after the new study as $\beta_{k} \mid D_1,D_2 \sim \N(\mu_{2k},\sigma_{2k}^{2})$. Here again the expected $\mu_{2k}$ can be assumed to equal $\mu_{1k}$. Approximating $\sigma_{2k}^{2}$ is a more difficult problem because of the study heterogeneity.  A simulation based solution is to generate results for the planned new study from the predictive distribution and update the meta-analysis with the new study \citepp{Sutton:evidencebased}.   
\citett{Roloff:planning} proposed a more straightforward approach to approximate $\sigma_{2k}^{2}$. Let $v$ denote the within-study variance in the new study which can be estimated on the basis of the within-study variances in the earlier studies. If the heterogeneity for the new study is assumed to be equal to the earlier studies, the variance after the new study can be approximated with
\begin{equation} \label{eq:randomeffect_sd}
 \sigma_{2k}^{2} = \sigma_{1k}^{2}  \frac{v+\gamma^2}{v+\gamma^2+\sigma_{1k}^{2}} .
\end{equation}
Ad hoc estimates for the overall heterogeneity exist for the cases where the new study is expected to decrease or increase heterogeneity \citepp{Roloff:planning}. Informative priors can be used in the cases where the heterogeneity cannot be reliably estimated because of the small number of existing studies \citepp{nikolakopoulou2015planning}.

To summarize, the following simplifying assumptions and approximations are made in the problem definition: the normality of the distributions, equality $\E(\mu_{2k})=\mu_{2k}$, approximation of $\sigma_{2k}$ and the assumption that $\gamma^2$ will be unchanged in the new study. The criteria that are presented in Section~\ref{sec:scientificimpact} are based on these assumptions and their numeric values are sensitive e.g. to the changes in $\sigma_{2k}$. However, it should be remembered that covariate prioritization essentially means ranking the covariates and the ranks are not likely to be very sensitive to the deviations from the assumptions.

\subsection{Covariate prioritization in meta-analysis with sparse selection priors} \label{sec:selectionpriors}
Above, it was assumed that the prior distribution is uninformative but in GWAS a viable alternative is to use sparse selection priors \citepp{george1993variable,o2009review} as informative prior distributions.
The sparseness means that with a high probability, the effect is exactly zero. For the non-zero effects, a weakly informative prior distribution is assumed. The collected data  update our information and the probability of a non-zero effect may increase. 

The spike-and-slab prior (with point-mass at zero) can be defined hierarchically
and independently for each effect $k$ as
\begin{equation*}
 \prod_k p(\beta_k | I_k) P( I_k) ,
\end{equation*}
where $P( I_k = 1) = \pi_{0}$ is a given small probability implying
that most effects are zero and non-zero effects are occurring only rarely. When data $D_1$ are collected the inclusion probabilities are updated to $P( I_k = 1 \mid  D_1 ) = \pi_{1k}$. After stage 2, the inclusion probabilities are further updated to $P( I_k = 1 \mid  D_1,D_2 ) = \pi_{2k}$.

Conditional distribution $p(\beta_k | I_k)$ is defined as follows
\begin{equation}  \label{eq:bayesI}
\begin{cases}
  \beta_k = 0, & \textrm{if } I_k=0, \\
  \beta_k \sim \N(\mu_{j},\sigma_{j}^{2}), & \textrm{if } I_k=1,
\end{cases}
\end{equation}
where $j=0$ before stage 1, $j=1$ after stage 1 and $j=2$ after stage 2.

Sparse selection priors can be seen as skeptical priors but in many cases they can be used present realistic prior odds. Sparse selection priors are applied to meta-analysis in the GWAS example considered in detail in Section~\ref{sec:illustration}. 

\section{Criteria for covariate prioritization} \label{sec:scientificimpact}
Next we will study several alternative criteria for the covariate prioritization by the realized and expected scientific impact. The criteria can be used to classify the covariates into categories I--III and to guide the decisions on the sample size. Out of the seven criteria presented, conditional power has been considered earlier \citepp{Roloff:planning,nikolakopoulou2014using} but the six other criteria have not been used earlier to measure scientific impact. The criteria represent different statistical principles which arise from frequentist, information theoretic and Bayesian paradigms. For non-Bayesian criteria, distributions~\eref{eq:prior} and \eref{eq:posterior}  can be interpreted as representing a point estimate and its asymptotic variance. The criteria are presented for the situation where the studies in stage 1 have been carried out and the studies in stage 2 are to be planned.  Without the loss of generality it is assumed that the regression coefficients are non-negative $\beta_k \geq 0$, $k=1,\ldots,K$. 

\subsection{Conditional power}
The idea of conditional power is closely related to hypothesis testing or more precisely to the Neyman-Pearson decision theory. The conditional power is defined as the power to reject the hypothesis $\beta_k=0$ after the new study is included in the meta-analysis. A benchmark hypothesis is defined as  $\beta_k=\delta_k$, where the smallest clinically significant effect size $\delta_k$  \citepp{copay2007understanding,revicki2008recommended} is a known constant.  In general, the clinically significant effect size varies covariate by covariate and depends on the measurement scale and the distribution of the covariate in the population \citepp{rankhazard}.

As a starting point, it is assumed that the meta-analysis of the earlier studies gives inconclusive results, i.e. the hypothesis $\beta_k=0$  cannot be rejected on the basis of the earlier studies. With significance level $\alpha$, the conditional power can be calculated as \citepp{Roloff:planning}
\begin{align} 
 \Phi \left( \frac{-C_{\alpha/2}\sigma_{2k}+\mu_{1k} \sigma_{1k}}{\sqrt{\frac{1}{\sigma_{2k}^2}-\frac{1}{\sigma_{1k}^2}}}  + \delta_k \sqrt{\frac{1}{\sigma_{2k}^2}-\frac{1}{\sigma_{1k}^2}}  \right) + \nonumber \\
 \Phi \left( \frac{-C_{\alpha/2}\sigma_{2k}-\mu_{1k} \sigma_{1k}}{\sqrt{\frac{1}{\sigma_{2k}^2}-\frac{1}{\sigma_{1k}^2}}}  - \delta_k \sqrt{\frac{1}{\sigma_{2k}^2}-\frac{1}{\sigma_{1k}^2}}  \right), \label{eq:condpower}
\end{align}
where $C_{\alpha/2}=\Phi^{-1}(1-\alpha/2)$ is the $100 (1-\alpha/2)$ percentile of the standard normal distribution (1.96 for $\alpha=0.05$). The conditional power measures the expected scientific impact in a straightforward manner: higher the power, higher the expected impact. In the Neyman-Pearson framework, the realized scientific impact is a binary variable: the hypothesis $\beta_k=0$ is either rejected or accepted after the new study.

The categorization presented in Introduction can be based on the conditional power. The covariates for which the hypothesis $\beta_k=0$ can be rejected already before the new study need not be studied further (category I). The division between category II and category III can be made by defining a selection limit in the terms of the conditional power. For instance, it could be required that the conditional power should be at least 80\%. The conditional power does not have a component for the importance of the covariate but it is implicitly assumed that all covariates are equally important regardless the value of $\beta_k$. 

\subsection{Change of p-value}
The use of p-values to prioritize covariates is closely related to Fisher's null hypothesis testing. It is well-known that the p-value of the null hypothesis $\beta_k=0$  does not alone measure the relevance of a covariate because the p-value can be made arbitrarily small by increasing the sample size. Therefore, the relevance must be decided comparing the effect size with an external benchmark that does not depend on the data but is based on expert knowledge.  Let $p_{1k}$ be the p-value related to the null hypothesis $\beta_k=\delta_k$ before the new study and $p_{2k}<p_{1k}$ be the p-value related to the same hypothesis after the new study. The realized scientific impact can then be defined as 
\begin{equation*}
 \log(p_{1k}) - \log(p_{2k}).
\end{equation*}
If $p_{2k}>p_{1k}$, the criterion is not applicable.
The expected scientific impact is obtained by  calculating the expected p-value after the new study
\begin{align} 
 & \log(p_{1k}) - \log(\E(p_{2k}))  = \nonumber \\
 & \log \left( 2 \left(1-\Phi \left( \frac{\mu_{1k}-\delta_k}{\sigma_{1k}} \right)  \right) \right) - \log \left( 2 \left(1-\Phi \left( \frac{\mu_{1k}-\delta_k}{\sigma_{2k}} \right) \right) \right), \label{eq:logp}
\end{align}
where $\Phi$ denotes the distribution function of the standard normal distribution. The condition $\E(p_{2k})<p_{1k}$ is fulfilled if $\mu_{1k}>\delta_k$.  The calculation utilizes the result $\E(\mu_{2k})=\mu_{1k}$ obtained from distribution~\eref{eq:prior}. Strictly speaking this can be seen as a violation against the frequentist principle that parameter $\beta_k$ is not random variable but a fixed unknown quantity.

Similarly to conditional power, covariates for which $p_{1k}$ is below a pre-specified limit can be classified into category I. The criterion gives a prioritization order for the rest of the covariates.  If $\mu_{1k} < \delta_k$, the criterion implies that the covariate is a member of category III and should not be studied further. This happens even if the current estimate is based on a very small sample.

\subsection{Change in lower confidence limit}
Confidence intervals quantify the uncertainty of an estimate and can be also used to measure the scientific impact. The main idea is to measure the distance from zero to the lower confidence limit. Assume that the confidence interval for $\hat{\beta}_k$ is $(l_{1k},u_{1k})$ before the new study and $(l_{2k},u_{2k})$ after the new study. The realized scientific impact can be measured by the change of the lower confidence limit
\begin{align*}
 & \max (0,l_{2k}) - \max (0,l_{1k}) = \\
 & \max \left( 0, \sigma_{2k}  \Phi^{-1}(\alpha/2) + \mu_{2k} \right) - \max \left( 0, \sigma_{1k} \Phi^{-1}(\alpha/2) + \mu_{1k}  \right).
\end{align*}
Utilizing the result $\E(\mu_{2k})=\mu_{1k}$, a criterion for the expected scientific impact can be obtained
\begin{align} 
 & \E(\max (0,l_{2k})) - \max (0,l_{1k}) = \nonumber \\
 & \max \left( 0, \sigma_{2k} \Phi^{-1}(\alpha/2) + \mu_{1k}  \right) - \max \left( 0, \sigma_{1k} \Phi^{-1}(\alpha/2) + \mu_{1k}  \right). \label{eq:lcl}
\end{align}
The covariates for which $l_{1k}>\delta_k$ can be classified into category I. The criterion ranks the rest of the covariates.

\subsection{Criterion based on Kullback-Leibler divergence}
Kullback-Leibler divergence \citepp{KullbackLeibler} is an information theoretic quantity for the asymmetric distance of two distributions. A straightforward attempt to quantify the scientific impact is to use the Kullback-Leibler divergence of the posterior from the prior\\
 $K(f_N(\mu_{2k},\sigma_{2k}^2) || f_N( \mu_{1k},\sigma_{1k}^2) )$,
which can be expressed in a closed form
\begin{align*}
 & K(f_N(\mu_{2k},\sigma_{2k}^2) || f_N( \mu_{1k},\sigma_{1k}^2) ) = \int_{-\infty}^{\infty} \log  \frac{f_N(x \mid \mu_{2k},\sigma_{2k}^2)}{f_N(x \mid \mu_{1k},\sigma_{1k}^2)}  f_N(x \mid \mu_{2k},\sigma_{2k}^2) \dd x =  \\
 & \frac{1}{2} \left( \frac{\sigma_{2k}^2}{\sigma_{1k}^2}+\frac{(\mu_{2k}-\mu_{1k})^2}{\sigma_{1k}^2}-1-\log  \frac{\sigma_{2k}^2}{\sigma_{1k}^2}  \right).
\end{align*}
The expected Kullback-Leibler divergence is then given by
\begin{align*}
 & \E [ K(f_N(\mu_{2k},\sigma_{2k}^2) || f_N( \mu_{1k},\sigma_{1k}^2) ) ] =  \\
 & \int_{0}^{\infty} \int_{-\infty}^{\infty} \frac{1}{2} \left( \frac{\sigma_{2k}^2}{\sigma_{1k}^2}+\frac{(\mu_{2k}-\mu_{1k})^2}{\sigma_{1k}^2}-1-\log \frac{\sigma_{2k}^2}{\sigma_{1k}^2}  \right) f(\mu_{2k},\sigma_{2k}^2 \mid \mu_{1k},\sigma_{1k}^2) \dd \mu_{2k} \dd \sigma_{2k}^2 =  \\
& -\frac{1}{2}-\frac{1}{2} \log  \frac{\sigma_{2k}^2}{\sigma_{1k}^2}  + \frac{\sigma_{2k}^2}{\sigma_{1k}^2}, 
\end{align*}
where the last equality follows from the assumption $\mu_{2k} \sim N(\mu_{1k},\sigma_{2k}^2)$ where $\sigma_{2k}^2$ is assumed to be known.
It can be seen that the expected Kullback-Leibler divergence depends only on the ratio of the variances. Therefore it is not a suitable criterion as such for the expected scientific impact because after the sufficient precision of the estimate has been achieved the additional decrease of variance does not have practical relevance. 

A more promising option is to use Kullback-Leibler divergence to quantify three components: the improvement of the precision $\sigma_{1k}^{2} \rightarrow \sigma_{2k}^{2}$, the change of the mean $\mu_{1k} \rightarrow \mu_{2k}$ and the clinical/practical importance of covariate $k$ as a predictor. The first two components describe the change in knowledge due to the new study and the third component describes the importance of the problem. In this approach, we first define an initial distribution $\N(0,\sigma_{Ik}^{2})$ for $\beta_k$ where the variance $\sigma_{Ik}^{2}$ is large to reflect the lack of information on $\beta_k$. Now the Kullback-Leibler divergence of the posterior from the initial distribution measures the importance of the covariate. The difference between the Kullback-Leibler divergence of the posterior from the initial distribution and the Kullback-Leibler divergence of the prior from the initial distribution measures the improvement of the precision and the change of the mean. To avoid the above mentioned problem of irrelevantly small variance, the variances can be modified by adding a positive term $\omega$ which bounds the importance of the variance. The realized scientific impact can be then defined as
\begin{align*}
  & K(f_N(\mu_{2k},\sigma_{2k}^2+\omega) || f_N(0 ,\sigma_{2k}^2+\omega) ) \cdot \nonumber \\ 
  & \left( K(f_N(\mu_{2k},\sigma_{2k}^2+\omega) || f_N( 0,\sigma_{Ik}^2+\omega) ) -  K(f_N(\mu_{1k},\sigma_{1k}^2+\omega) || f_N( 0,\sigma_{Ik}^2+\omega) ) \right). 
\end{align*}
The expected scientific impact can be approximated with 
\begin{align} 
 & \E(K(f_N(\mu_{2k},\sigma_{2k}^2+\omega) || f_N( 0,\sigma_{2k}^2+\omega) )) \cdot \nonumber \\
 & \left[ \E(K(f_N(\mu_{2k},\sigma_{2k}^2+\omega) || f_N(0,\sigma_{Ik}^2+\omega) )) -  K(f_N(\mu_{1k},\sigma_{1k}^2+\omega) || f_N( 0,\sigma_{Ik}^2+\omega) ) \right] = \nonumber \\
& \frac{1}{4} \left( \frac{\mu_{1k}^2+\sigma_{1k}^2}{\sigma_{1k}^2+\omega}\right) \left( \frac{2 \sigma_{1k}^2-\sigma_{1k}^2}{\sigma_{Ik}^2+\omega}-\log \frac{\sigma_{1k}^2+\omega}{\sigma_{1k}^2+\omega}\right). \label{eq:KL}
\end{align}
The criterion ranks the covariates but does not give direct indication whether a covariate belongs to category I. Some other criterion can be used first to identify covariates that belong to category~I.

\subsection{Difference between prior and posterior expectation} \label{subsec:bayes}
Bayesian criteria can be defined on the basis of the model with uninformative priors~\eref{eq:posterior} or the model with the sparse selection priors~\eref{eq:bayesI}. We concentrate on the latter case where the most of the effects are zero. This is often a reasonable assumption in genetics and other fields where a massive number of potential covariates are tested.

When planning a new study, the posterior $p(\beta_k \vert D_1, D_2)$  is not known. The expected posterior can be approximated by
\begin{align}
 & \hat{\sigma}_{2k}^{2} = \frac{1}{\frac{1}{\sigma_{1k}^{2}}+\frac{1}{\hat{\sigma}_{Dk}^{2}}} \nonumber \\
 & \hat{\mu}_{2k}= \hat{\sigma}_{2k}^{2} \left(  \frac{\mu_{1k}}{\sigma_{1k}^{2}} + \frac{\hat{\mu}_{Dk}}{\hat{\sigma}_{Dk}^{2}}   \right)  \label{eq:expected_posterior} \\
 & \hat{\pi}_{2k} = \frac{\pi_{1k} f_N(\hat{\mu_{Dk}} \mid 0,\hat{\sigma}_{Dk}^{2})}{\pi_{1k} f_N(\hat{\mu}_{Dk} \mid 0,\hat{\sigma}_{Dk}^{2} )+(1-\pi_{1k})f_N(\hat{\mu}_{Dk} \mid \hat{\mu}_{2k} ,\hat{\sigma}_{Dk}^{2})} \nonumber,
\end{align}
where mean $\hat{\mu}_{Dk}$ and variance $\hat{\sigma}_{Dk}^{2}$ are related to the new study to be conducted. Mean $\hat{\mu}_{Dk}$ is taken to be equal to the mean of the current data and variance $\hat{\sigma}_{Dk}^{2}$ is approximated on the basis of the current data and the sample size of the new study.   

We propose three Bayesian criteria to measure expected scientific impact. The first criterion uses the difference between prior and posterior expectation
\begin{equation} \label{eq:prior_posterior_diff}
 \hat{\pi}_{2k} \hat{\mu}_{2k} - \pi_{1k} \mu_{1k}
\end{equation}
as the measure. Here $\mu_{1k}$ and $\hat{\mu}_{2k}$ measure the importance of the problem and $\pi_{1k}$ and $\hat{\pi}_{2k}$ measure the precision. The criterion ranks the covariates but does not give direct indication if a covariate with a low value of the criterion belongs to category I or III. This can be concluded using, for instance, Bayes factors presented next.

\subsection{Change in Bayes factor}
Commonly used Bayes factors (BF) have attractive properties as measures of evidence and can be also utilized in measuring scientific impact.
The evidence is combined from multiple data sources \citepp{Ball2007,Wakefield2008} and  
similarly to likelihood ratio statistic compared against a null model \citepp{Lee2000}. For each $k$, BF is calculated as the posterior inclusion ratio divided by the prior inclusion ratio 
\begin{equation} \label{eq:BF}
 \textrm{BF}_k = \frac{P(I_k=1 \mid D_1,D_2)}{P(I_k=0 \mid D_1,D_2)} \Big/ \frac{P(I_k=1 \mid D_1)}{P(I_k=0 \mid D_1)}.
\end{equation}
\citett{Jeffreys:theoryofprobability} have provided following guidelines to make a formal decision based on BFs between: 1 and 3
(barely worth mentioning), 3 and 10 (substantial support), 10 and 100 (strong support), more than 100 (decisive support). With sparse selection priors, the BF limits can be significantly larger. 

BFs are useful in the interpretation of ‘significant’ associations \citepp{Ioannidis2008} because the interpretation is independent from the sample size (unlike p-values) \citepp{Wakefield2009}. However, assume that we already have collected lot of data and the prior probability $P(I_k=1 \mid D_1)$ is close to one for given $k$. Now, the criterion for scientific impact should tell that collecting more data may not change our knowledge. The direct usage of BF fails to do this: when the posterior $P(I_k=0 \mid D_1, D_2)$ approaches to zero, the criterion increases without bounds. Therefore, scientific impact should  be measured comparing BF before and after the new study. A covariate belongs to category I if the BF before the new study is greater than the limit of decisive support and to category III the expected BF after the new study is below this limit. If the new study is expected to increase the BF so that it will exceed the limit, the covariate belongs to category II. 

\subsection{Criterion based on Bayesian false discovery rate}
The third Bayesian criterion utilizes Bayesian false discovery rate (BFDR) control approach \citepp{ventrucci2011multiple,efron2001empirical} separately for the distributions before and after the new study. When the posteriors are considered, we approximate the local false discovery rate (LFDR) for each covariate $k$ with the probability   $P(I_k=0 \mid D_1,D_2)$. Global level BFDR can then be calculated (at given probability level) by combining these local quantities over the covariates. This is done by ordering $\textrm{LFDR}_k=P(I_k=0 \mid D_1,D_2)$ values to the ascending order and averaging the smallest $\textrm{LFDR}_k$ values until the average is still smaller than given probability level (e.g., 0.05). All covariates contributing to the average are then declared to be `significant' or `interesting'. Note that BFDR also contains multiplicity adjustment \citepp{scott2010bayes}. When the distributions before the new study are considered, we repeat the same procedure with the probabilities $P(I_k=0 \vert D_1)$. As an outcome, we have two sets of covariates: one from the BFDR analysis before the new study and one from the BFDR analysis after the new study. The covariates in the latter set but not in the former set are considered as the covariates in category II.

\section{Illustration: GWAS for CRP} \label{sec:illustration}

In the illustrative example, the criteria presented in Section~\ref{sec:scientificimpact} are applied to guide the design of new studies on genetic variants associated with CRP. The source code to reproduce the results is available at\\ \url{http://www.tilastotiede.fi/papers/Karvanen_Sillanpaa_GWAS_example.zip}. As a starting point, we consider a fixed-effects meta-analysis of GWAS in over 80,000 subjects \citepp{dehghan2011meta}. In this meta-analysis, 17 loci were identified in a ``discovery panel'' of 15 studies and 66,185 participants. The identified 17 loci were studied further in a ``replication panel'' of 10 studies and 16,540 participants. The 17 loci are listed in Table~\ref{tab:GWAS} together with estimates from the replication panel. Evidence of replication was provided for seven loci \citepp{dehghan2011meta}. In other words, there were 10 loci for which the evidence of replication could not be provided. We use the example only to demonstrate the covariate prioritization and do not discuss general issues related to the analysis of GWAS replications \citepp{ioannidis2001replication,skol2006joint,sillanpaa2004replication}.

Now the question of interest is the selection of loci for a new replication. For this, we rank the loci according to the expected scientific impact measured by the seven criteria presented in Section~\ref{sec:scientificimpact}. The new replication is assumed to share the essential characteristics with the original replication and to have the same sample size. \citett{dehghan2011meta} found that the between-study heterogeneity was statistically significant in their meta-analysis but they do not report results from a random-effects meta-analysis. Without the estimated between-study variances, we are compelled to follow their example and to base our calculations on a fixed-effects meta-analysis.
In the illustration, we restrict to 17 loci in Table~\ref{tab:GWAS} although in practice it would be reasonable to also consider  other loci from GWAS when designing a new replication. 

Parameter $\beta_k$ describes the effect of SNP $k$ on the log-CRP in a linear model  \citepp{dehghan2011meta}. The user defined parameters for the criteria were set as follows. On the basis of the results on the association of CRP and cardiovascular diseases \citepp{danesh2000low,pepys2003c}, we set clinically significant effect size for log-CRP to be $\delta_k=0.03$. This value was used for the change in p-value and conditional power.  Following the original report \citepp{dehghan2011meta}, significance level $\alpha=6.9\times10^{-4}$ was selected for conditional power and for the change in lower confidence level.  Large initial variance $\sigma_{Ik}^2=100$ and small variance limit $\omega=0.01^2$ were used for Kullback-Leibler divergence based criterion. For the Bayesian criteria, we used the sparse selection priors presented in Section~\ref{sec:selectionpriors}. We expected one SNP out of a million to be associated with CRP and used prior inclusion probability $\pi_{0}=10^{-6}$ for all SNPs. Consistently with this prior, a limit of one million is used for BF. BFDR we chose to be $0.05$. 

The results are presented in Table~\ref{tab:GWAS}. The genes are ordered by their p-value in the discovery panel. It is investigated which  genes have the highest expected scientific impact according to different criteria. It is seen that CRP, APOC1 and HNF1A can be classified into category I by both frequentist and Bayesian criteria. All the criteria rank the genes LEPR, IL6R and IL1F10 among the four top genes. The conditional power is high also for GCKR. Change in p-value, change in lower confidence limit and Kullback-Leibler divergence based criterion   identify also SALL1 as a gene to be studied further. Overall, the results are quite consistent, even between the frequentist and the Bayesian criteria.

 \begin{landscape}
 \begin{table}
 \caption{Application of covariate prioritization in design of a new replication study on genetic variants associated with CRP. Effect sizes estimated from the discovery panel and from the replication panel are presented for the 17 loci that were associated with CRP in the discovery panel \citepp{dehghan2011meta}. Estimated $\beta$ represents one-unit change in the natural log-transformed CRP (mg/L) per copy increment in the coded allele. The frequentist criteria measure the  expected scientific impact for the new replication by conditional power (CP), change of p-value ($\Delta$log(p)), change of lower confidence limit (LCL) and Kullback-Leibler (KL) criterion~\eref{eq:KL} multiplied by 1000. $\Delta \log(p)$ would be negative if $\beta<\delta_k$ and is not reported. A result with superscript $^*$ indicates that the gene belongs to category I. The Bayesian criteria include difference between prior and posterior expectation~\eref{eq:prior_posterior_diff} $\Delta\E(\beta)$ and BF and BFDR analyses where 1--1 indicates category I, 1--0 indicates category II and 0--0 indicates category III. The highest expected scientific impacts according to each criteria are bolded.
} 
\label{tab:GWAS}
\begin{longtable}{llcc|cccc|ccc} 
Gene & SNP    & {Discovery} & {Replication} & \multicolumn{4}{c}{Frequentist criteria} & \multicolumn{3}{c}{Bayesian criteria}  \\
 & (coded allele) &  $\beta$ (SE) & $\beta$ (SE) & CP & $\Delta$log(p)  & LCL  & KL & $\Delta\E(\beta)$ & BF  & BFDR \\
  CRP  &  rs2794520 (C)  &  0.193 (0.007)  &  0.086 (0.010)  &  1.00$^*$  &  16.0$^*$  &  0.010$^*$  &  3570$^*$  &  0.000$^*$  &  1--1  &  1--1 \\ 
 APOC1  &  rs4420638 (A)  &  0.240 (0.010)  &  0.200 (0.032)  &  0.99$^*$  &  14.4$^*$  &  0.032$^*$  &  10060$^*$  &  0.001$^*$  &  1--1  &  1--1 \\ 
 HNF1A  &  rs1183910 (G)  &  0.152 (0.007)  &  0.122 (0.021)  &  0.99$^*$  &  9.9$^*$  &  0.021$^*$  &  6168$^*$  &  0.005$^*$  &  1--1  &  1--1 \\ 
 LEPR  &  rs4420065 (C)  &  0.111 (0.007)  &  0.045 (0.009)  &  \textbf{1.00}  &  \textbf{1.6}  &  \textbf{0.009}  &  \textbf{931}  &  \textbf{0.035}  &  \textbf{1--0}  &  \textbf{1--0} \\ 
 IL6R  &  rs4129267 (C)  &  0.094 (0.007)  &  0.045 (0.010)  &  \textbf{1.00}  &  \textbf{1.4}  &  \textbf{0.010}  &  \textbf{995}  &  \textbf{0.044}  &  \textbf{1--0}  &  \textbf{1--0} \\ 
 GCKR  &  rs1260326 (T)  &  0.089 (0.007)  &  0.031 (0.010)  &  \textbf{0.90}  &  0.0  &  0.007  &  485  &  0.000  &  0--0  &  0--0 \\ 
 NLRP3  &  rs12239046 (C)  &  0.048 (0.007)  &  0.042 (0.018)  &  0.21  &  0.4  &  0.000  &  885  &  0.000  &  0--0  &  0--0 \\ 
 IL1F10  &  rs6734238 (G)  &  0.047 (0.007)  &  0.072 (0.017)  &  \textbf{0.89}  &  \textbf{3.3}  &  \textbf{0.017}  &  \textbf{2530}  &  \textbf{0.070}  &  \textbf{1--0}  &  \textbf{1--0} \\ 
 PPP1R3B  &  rs9987289 (G)  &  0.079 (0.011)  &  0.003 (0.031)  &  0.00  &  --  &  0.000  &  127  &  0.000  &  0--0  &  0--0 \\ 
 ASCL1  &  rs10745954 (A)  &  0.043 (0.006)  &  0.018 (0.015)  &  0.05  &  --  &  0.000  &  218  &  0.000  &  0--0  &  0--0 \\ 
 HNF4A  &  rs1800961 (C)  &  0.120 (0.018)  &  0.023 (0.026)  &  0.00  &  --  &  0.000  &  283  &  0.000  &  0--0  &  0--0 \\ 
 RORA  &  rs340029 (T)  &  0.044 (0.007)  &  0.004 (0.010)  &  0.08  &  --  &  0.000  &  32  &  0.000  &  0--0  &  0--0 \\ 
 SALL1  &  rs10521222 (C)  &  0.110 (0.017)  &  0.089 (0.028)  &  0.29  &  \textbf{2.5}  &  \textbf{0.022}  &  \textbf{2475}  &  \textbf{0.002}  &  0--0  &  0--0 \\ 
 PABPC4  &  rs12037222 (A)  &  0.047 (0.008)  &  0.035 (0.017)  &  0.16  &  0.1  &  0.000  &  650  &  0.000  &  0--0  &  0--0 \\ 
 BCL7B  &  rs13233571 (C)  &  0.054 (0.010)  &  0.049 (0.025)  &  0.05  &  0.5  &  0.000  &  927  &  0.000  &  0--0  &  0--0 \\ 
 PSMG1  &  rs2836878 (G)  &  0.040 (0.007)  &  0.013 (0.011)  &  0.19  &  --  &  0.000  &  114  &  0.000  &  0--0  &  0--0 \\ 
 RGS6  &  rs4903031 (G)  &  0.046 (0.008)  &  0.001 (0.012)  &  0.01  &  --  &  0.000  &  37  &  0.000  &  0--0  &  0--0 
  \end{longtable}
\end{table}
\end{landscape} 

\begin{figure}
\begin{center}
\includegraphics[width=0.95\columnwidth]{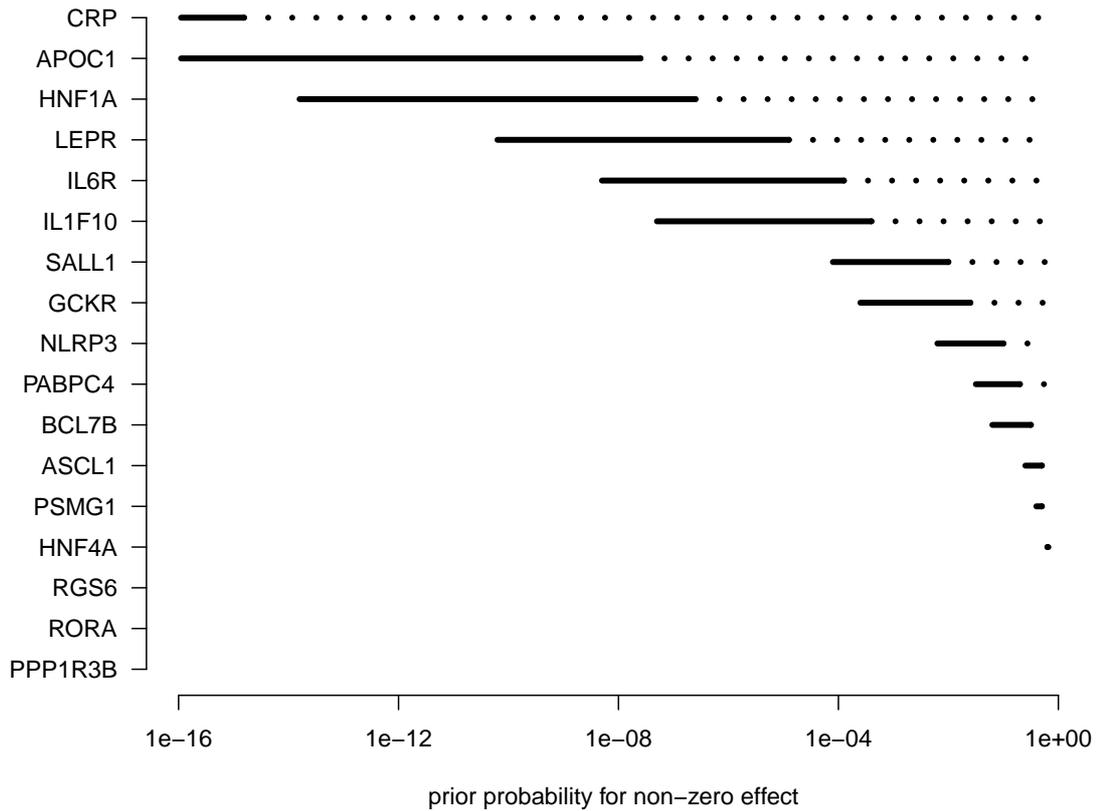}
\caption{Sensitivity of BFDR analysis as a function of prior probability of non-zero effect size. The solid line indicates that the SNP should be included in the new replication analysis for the given prior probability. The dotted line indicates that it is not necessary to include the SNP in the new replication analysis because the existing results are already conclusive. The absence of a line indicates that the gene should not be included in the replication because it is unlikely that a non-zero posterior effect could be found in the meta-analysis after the new study.    \label{fig:priorsensitivity}}
\end{center}
\end{figure}

The sensitivity to the choice of user defined parameters is studied in the Bayesian setup where the prior probability of non-zero effect size is set to have a wide range of values from $10^{-16}$ to $10^{-0.2}$. The results based on the BFDR analysis under these settings are given in Figure~\ref{fig:priorsensitivity}. It can be seen that the selection of SNPs strongly depends on the prior probability but there is a clear logic behind the results. When the prior probability is small, most of SNPs are not selected because the additional data would not be sufficient to change the conclusions (category III). When the prior probability increases, the new study is expected to change the prior non-significance to posterior significance (category II) and SNP becomes selected. When the prior probability further increases, the evidence is sufficient even without the new study (category I) and the SNP is not selected. Naturally, the prior must be the same for both the design and the analysis. Thus, the question of the prior selection is not specific for the covariate prioritization but common for Bayesian analysis in general.

\begin{figure}
\begin{center}
\includegraphics[width=0.95\columnwidth]{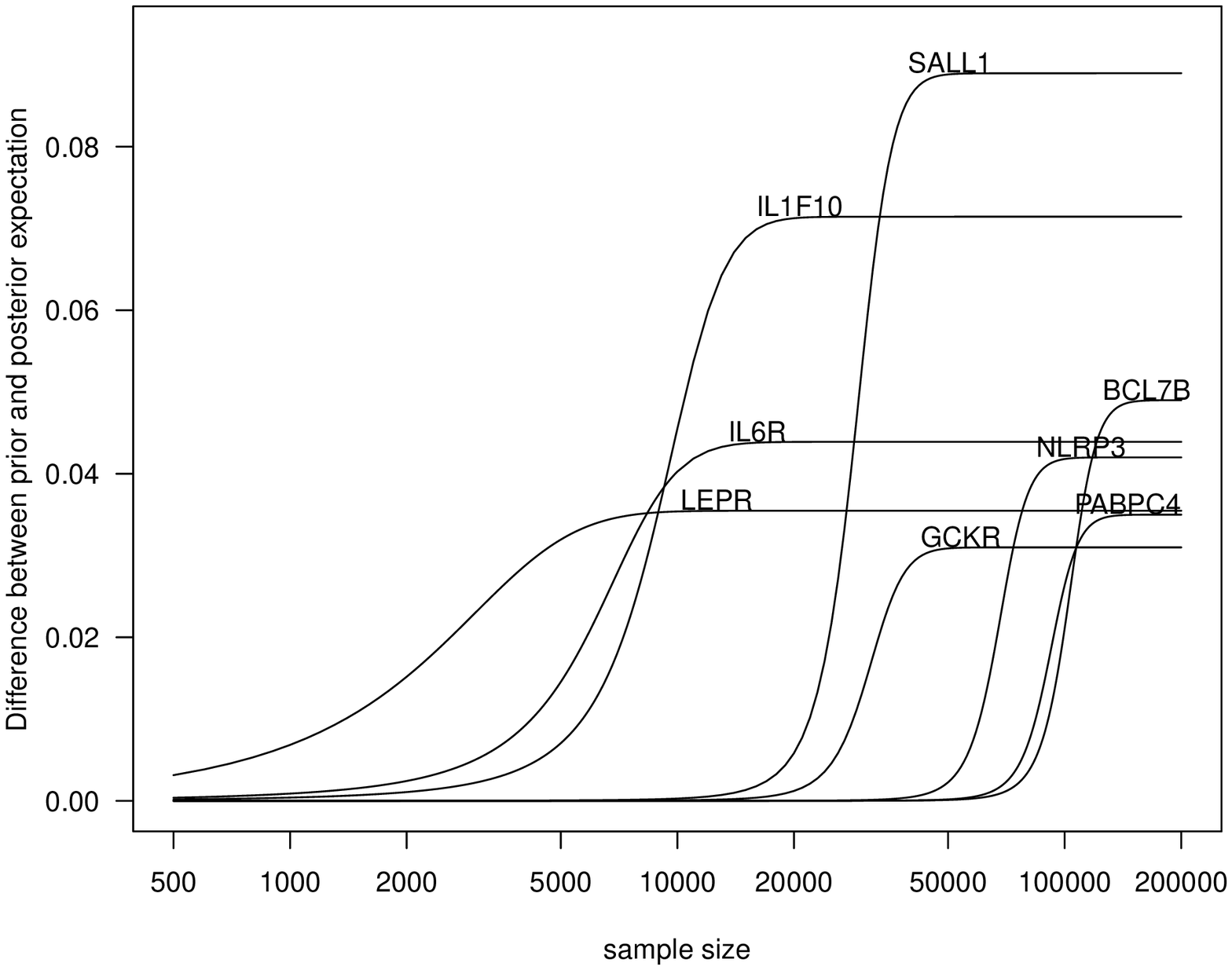}
\caption{The impact of sample size of the new replication to expected scientific impact measured by $\Delta\E(\beta)$. The prior inclusion probability was set at $\pi_{00}=10^{-6}$. The genes for which $\Delta\E(\beta)<0.01$ even for sample size 200,000 are not displayed.  \label{fig:samplesize}}
\end{center}
\end{figure}

To illustrate how the criteria can be used to guide the decisions on the sample size we study the expected scientific impact as a function of the sample size of the replication study. We measure the expected scientific impact by the difference between prior and posterior expectation but the other criteria could be applied as well. The results in Figure~\ref{fig:samplesize} can be used to conclude the sample size required to have a desired expected scientific impact. For instance, the sample sizes below 14,000 seem to be  insufficient for gene SALL1. Increasing the sample size after a certain limit does not bring benefits. For instance, increasing the sample size from 10,000 to 200,000 does not increase the expected scientific impact for LEPR.  The graph also tells that the gene with the highest expected scientific impact is LEPR when the sample size is below 8300, then become IL6R and ILF10 and finally SALL1 when the sample size is above 33,000. These genes are the same as the top genes in Table~\ref{tab:GWAS}.
 
\section{Discussion} \label{sec:discussion}
We have proposed that covariate prioritization should be based on a formal decision criterion
and have introduced a variety of new criteria to aid the study design in situations where
there already is previous knowledge. The prioritization is based on the idea that the covariates in the category II should be preferred in replication studies. The criteria can be also used to guide the decisions on the sample size. The example on GWAS replication demonstrates the applicability of the proposed approaches in covariate prioritization and sample size determination. 

Each criterion defined in Section~\ref{sec:scientificimpact} requires some parameters to be specified. Guidelines exist for the specification of clinically significant effect size $\delta_k$ \citepp{copay2007understanding,revicki2008recommended} as well as for the elicitation of experts' opinions about prior distributions \citepp{ohagan:eliciting}. The selection by a formal criterion encourages the researcher to express his or her preferences explicitly in a quantitative form. In addition to the presented criteria, other alternatives can be formulated. For instance, frequentist false discovery rate \citepp{benjamini1995controlling} could be used in a similar manner as BFDR in Section~\ref{subsec:bayes}. Likewise, the overlapping coefficient \citepp{inman1989overlapping} and Hellinger distance \citepp{pollard2002user} used by \citett{nikolakopoulou2015planning} could be also applied to covariate selection.

In some cases, the presented criteria may lead to different choices of covariates to be studied. We recommend the researchers to use a criterion that is in agreement with the statistical analysis to be carried out. For instance, conditional power is closely related to the Neyman-Pearson type decision making and the Bayesian criteria are natural choices with a Bayesian analysis. Sensitivity analyses similar to one in Figure~\ref{fig:priorsensitivity} are recommended for learning about the impact of the user defined parameters to the prioritization. 

In the presented example, it was realistic to assume that the marginal costs of measurement are the same for all covariates. This assumption can be relaxed by first calculating the affordable sample size for each covariate and then applying the formal criterion with these sample sizes. It is also possible to extend the covariate prioritization and sample size determination for the design of multiple studies.

In general, covariate prioritization using a formal criterion is expected to work best in scenarios such as GWAS where a large number of covariates are used independently from each other in the analysis. The approach may not be useful in scenarios the data are modeled as a complex causal network between variables and confounding is a major issue. 

We hope that this work will encourage researchers to consider the study design in a systematic way. As a benefit, researchers would be able to make smarter decisions on the allocation of their time, effort and resources.

~\\
\noindent {\bf{Acknowledgement}} The authors are grateful to Georgia Salanti for many helpful comments that significantly improved the manuscript.



\bibliographystyle{apalike}
\bibliography{impact}

\end{document}